\documentstyle[12pt,iopconf]{article}
\begin{document}
\title{Global statistical model calculations and the role of isospin}

\author{Thomas Rauscher\footnote{APART fellow of the Austrian Academy of
Sciences}
and Friedrich--Karl Thielemann}

\affil{Institut f\"ur Physik, University of Basel, Klingelbergstr.\ 82,
CH-4056 Basel, Switzerland}

\beginabstract
An improved code for the calculation of astrophysical reaction rates in
the statistical model is presented. It includes the possibility to study
isospin effects. Such effects heavily affect rates involving
self--conjugate nuclei and may also be found in reactions on other
intermediate and heavy targets.
\endabstract

\section{Introduction}

The investigation of explosive nuclear burning in astrophysical
environments is a challenge for both theoretical and experimental
nuclear
physicists. Highly unstable nuclei are produced in such processes which
again can be targets for subsequent reactions. Cross sections and
astrophysical reaction rates for a large number of nuclei
are required to perform complete network
calculations which take into account all possible reaction links and do
not postulate a priori simplifications.

The majority of reactions can be described in the framework of the
statistical model (compound nucleus mechanism, Hauser--Feshbach
approach, HF)~\cite{hau52}, provided that the level density of the
compound nucleus
is sufficiently large in the contributing energy window~\cite{rau97}.
In astrophysical applications usually different aspects are emphasized
than in pure nuclear physics investigations. Many of
the latter in this long and well established field were focused on
specific reactions, where all or most "ingredients", like optical
potentials for
particle transmission coefficients, level densities, resonance
energies and widths of giant resonances to be implemented in
predicting E1 and M1 $\gamma$--transitions, were deduced from
experiments.
As long as the statistical model prerequisites are met, this will
produce
highly accurate cross sections.
For the majority of nuclei in astrophysical applications such
information is not available. The real challenge is thus not the
well--established statistical model, but rather to provide all these
necessary ingredients
in as reliable a way as possible, also for nuclei where none of such
information is available.

In Section~\ref{nosmo}, an improved code for the calculation of
astrophysical reaction rates in the statistical model is briefly
presented. It includes the possibility of studying isospin effects. The
latter are further discussed in Section~\ref{isosec}.

\section{The code NON--SMOKER\label{nosmo}}

Based on the well--known code SMOKER~\cite{thi87}, an improved code for
the prediction of astrophysical cross sections and reaction rates in the
statistical model has
been developed. The current status of the new code NON--SMOKER is
described in the following.

The final quantities entering the expression for the cross
section in the statistical model~\cite{hau52} are the averaged
transmission
coefficients. They do not reflect a resonance behavior but rather
describe absorption via an imaginary part in the (optical)
nucleon--nucleus potential~\cite{mah79}. In astrophysics, usually
reactions induced by light projectiles (neutrons, protons, $\alpha$
particles) are
most important. Global optical potentials are quite well defined for
neutrons and protons. It was shown~\cite{thi87,thi83,cow91} that the best fit
of s--wave neutron strength functions is obtained with the optical
potential by~\cite{jeu77}, based on microscopic
infinite nuclear matter calculations for a given density,
applied with a local density approximation.
It includes corrections of the imaginary part~\cite{fantoni81,mahaux82}.
A similar description is used for protons.

Optical potentials for $\alpha$ particles are treated in the folding
approach~\cite{sat79}, with a parametrized mass-- and energy--dependence
of the real volume integral~\cite{rau98}. The mass-- and
energy--dependence of the imaginary potential is parametrized according
to~\cite{bro81} and additionally includes microscopic and deformation
information~\cite{rau98}.

Deformed nuclei are treated by an effective spherical potential of equal
volume~\cite{thi87,cow91}.
For a detailed description of the formalism used to calculate E1 and M1
$\gamma$--transmission coefficients and the inclusion of width fluctuation
corrections~\cite{tep74}, see~\cite{rau97} and references therein.

The level density treatment has been recently improved~\cite{rau97}.
However, the problem of the parity distribution at low energies remained.
The new code includes a modified version of the
description~\cite{rau97}, accounting for non--evenly distributed
parities at low energies, based on most recent findings within the
framework of the shell model Monte Carlo method~\cite{alh97}.

Additionally, the included data set of experimental level information
(excitation energies, spins, parities) has been updated~\cite{ENSDF},
as well as the experimental nuclear masses~\cite{audi}. These data bases
are continuously updated. For theoretical masses, there is a choice
between different mass models (e.g.\ by Hilf \etal~\cite{hilf},
FRDM~\cite{moeller}, ETFSI~\cite{abu}), of which currently the FRDM
is favored. Microscopic information needed for the calculation of level
densities and $\alpha$+nucleus potentials are also taken from the FRDM, as
well as experimentally unknown ground state spins~\cite{moellkra}.

Finally, isobaric analog states $T^>=T^<+1=T^{\rm g.s.}+1$ are
explicitly considered in the new code. This will be discussed in the
next section.

\section{Inclusion of isospin\label{isosec}}

The original HF equation~\cite{hau52} implicitly assumes complete
isospin mixing but can be generalized to explicitly treat the
contributions of the dense background states with isospin $T^<=T^{\rm
g.s.}$ and the isobaric analog states with
$T^>=T^<+1$~\cite{gri71,har77,sar82,har86}. In reality, compound nucleus
states do not have unique isospin and for that reason an isospin mixing
parameter $\mu\downarrow$ was introduced~\cite{gri71}, which is the
fraction of the width of $T^>$ states leading to $T^<$ transitions.
For complete isospin mixing $\mu\downarrow=1$, for pure $T^<$ states
$\mu\downarrow=0$. In the case of overlapping resonances for each
involved isospin, $\mu\downarrow$ is directly related to the level
densities $\rho^<$ and $\rho^>$, respectively. Isolated resonances can
also be included via their internal spreading width
$\Gamma^{\downarrow}$ and a bridging formula was derived to cover both
regimes~\cite{lan78}.

In order to determine the mixing parameter
$\mu\downarrow=\mu\downarrow(E)$, experimental information for
excitation energies of $T^>$ levels is used where available~\cite{ENSDF,rei90}
in the code NON--SMOKER.
Experimental values for spreading widths are also
tabulated~\cite{har86,rei90}. Similarly to the standard treatment for
the $T^<$ states, a level density description~\cite{rau97} is invoked 
above the last experimentally known
$T^>$ level. Since the $T^>$ states in a
nucleus ($Z$,$N$) are part of 
multiplet, they can be approximated by the levels (and
level density) of the nucleus ($Z$$-$1,$N$+1), only shifted by a certain
energy $E_{\rm d}$. This displacement energy can be
calculated~\cite{aue72} and it is dominated by the Coulomb displacement
energy $E_{\rm d}=E_{\rm d}^{\rm Coul}+\epsilon$.

The inclusion of the explicit treatment of isospin has two major effects
on statistical cross section calculations in astrophysics: 
the suppression of $\gamma$--widths for reactions involving
self--conjugate nuclei and the suppression of the neutron emission
in proton--induced reactions. Non--statistical effects, i.e.\ the
appearance of isobaric analog resonances, are included in the treatment
of the mixing parameter $\mu\downarrow$~\cite{lan78} but will not be
further discussed here.

\subsection{$\gamma$--Widths}

The isospin selection rule for $E1$ transitions is $\Delta T=0,1$ with
transitions $0\rightarrow0$ being forbidden~\cite{jon87}. An approximate
suppression rule for $\Delta T=0$ transitions in self--conjugate nuclei
can also be derived for $M1$ transitions~\cite{jon87}.

In the case of ($\alpha$,$\gamma$) reactions on targets with $N=Z$, the
cross sections will be heavily suppressed because $T=1$ states cannot be
populated due to isospin conservation. A suppression will also be found
for capture reactions leading into self--conjugate nuclei, although
somewhat less pronounced because $T=1$ states can be populated according
to the isospin coupling coefficients.

In previous reaction rate calculations~\cite{woo78,cow91,sch98} the
suppression of the $\gamma$--widths was treated completely
phenomenologically by dividing by quite uncertain factors of 5 and 2, for
($\alpha$,$\gamma$) reactions and nucleon capture reactions,
respectively. In the new code NON--SMOKER, the appropriate
$\gamma$--widths are automatically obtained, by explicitly accounting for
population and decay of $T^<$ and $T^>$ states, and considering isospin mixing
by the parameter $\mu\downarrow$.

\subsection{Competition cusps in proton--induced reactions}

Assuming incomplete isospin mixing, the strength of the neutron channel
will be suppressed in comparison to the proton channel in 
reactions p+target~\cite{gri71,sar82,har86}. This leads to a smaller
cross section for (p,n) reactions and an increase in the cross section
of (p,$\gamma$) reactions above the neutron threshold, as compared to
calculations neglecting isospin (i.e.\ implicitly assuming complete
isospin mixing with $\mu\downarrow=1$).

The isospin mixing parameter was varied in the theoretical investigation
of a $^{51}$V(p,$\gamma$)$^{52}$Cr experiment~\cite{zys80}. It was 
found~\cite{zys80} that complete isospin mixing closely reproduced the
measured cross sections when width fluctuation corrections were
considered. Width fluctuation corrections~\cite{ver84} 
affect the (p,$\gamma$) cross
sections above as well as below the neutron threshold, whereas
incomplete isospin mixing only reduces the cross sections above the
threshold. Thus, the two corrections can be discriminated. Mainly from
that result, it was concluded that --- contrary to width fluctuation
corrections --- isospin can be neglected.

However, a closer investigation of the $T^>$ levels in $^{52}$Cr 
(using \cite{lan78} and \cite{rau97}) shows that
isospin mixing should be rather complete already at the neutron threshold
(since the first $T^>$ state is almost 1 MeV below the
threshold~\cite{ENSDF}). This
is also true for lighter targets. Nevertheless, for reactions on more
heavy nuclei ($Z>30$) the neutron and proton threshold, respectively,
will still be in a region of
incomplete isospin mixing and therefore isospin effects should be
detectable there. This effect, however, does not play such an important
role in the calculation of astrophysical reaction rates as the
suppression of the $\gamma$--width described in the previous chapter,
because of the averaging over an energy range (the Gamow window) in the
calculation of the rate.

\section{Conclusion}

The new code NON--SMOKER makes use of the latest set of descriptions for the
calculation of the nuclear properties needed to reliably predict
astrophysical reaction rates, such as masses, level densities, nucleon--
and $\alpha$--potentials, GDR energies and widths, width fluctuation
corrections. Additionally, the possibility of studying
isospin effects has been included. This also leads to a more fundamental
treatment of the $\gamma$--width suppression for compound nuclei with
$T^{\rm g.s.}=T^<=0$.

Nevertheless, more experimental data are needed to check and further improve 
current parametrizations.
Especially investigations over a large mass range would
prove useful to fill in gaps in the knowledge of the nuclear structure
of many isotopes and to construct more powerful parameter systematics.
Such investigations should include neutron--,
proton-- and $\alpha$--strength functions, as well as radiative widths, and
charged particle scattering and reaction cross sections for {\it stable}
and unstable isotopes. More capture data with
self--conjugate final nuclei would also be highly desireable.

This information can be used to make future
large--scale statistical model calculations even more accurate.

\end{document}